\documentclass[twocolumn,preprintnumbers,amsmath,amssymb]{revtex4}
\usepackage{psfrag}
\usepackage{amssymb,amsmath,amsthm}
\usepackage[dvips]{graphicx}
\usepackage{verbatim}
\usepackage{graphicx}
\usepackage{dcolumn}% Align table columns on decimal point
\usepackage{bm}% bold math

\begin{document}

%\title{SPATIOTEMPORAL PHASE SYNCHRONIZATION IN A LARGE ARRAY OF CONVECTIVE OSCILLATORS}
\title{Spatiotemporal phase synchronization in a large array of convective oscillators}
%\author{M. A. MIRANDA and J. BURGUETE}
%\email[e-mail address:\ ]{montse@alumni.unav.es}
%\affiliation{Dept. of Physics and Applied Mathematics, Universidad de Navarra. Irunlarrea s/n, E-31080 Pamplona, Spain}
\author{M. A. Miranda}
\email[e-mail address:\ ]{montse@alumni.unav.es}
\affiliation{Dept. of Physics and Applied Mathematics, Universidad de Navarra. Irunlarrea s/n, E-31080 Pamplona, Spain}

\author{J. Burguete}
\email[e-mail address:\ ]{javier@fisica.unav.es}
\affiliation{Dept. of Physics and Applied Mathematics, Universidad de Navarra. Irunlarrea s/n, E-31080 Pamplona, Spain}

\begin{abstract}

\noindent In a quasi-1D thermal convective system consisting of a large array of nonlinearly coupled oscillators, clustering is the way to achieve a regime of mostly antiphase synchronized oscillators. This regime is characterized by a spatiotemporal doubling of traveling modes. As the dynamics is explored beyond a spatiotemporal chaos regime (STC) with weak coupling, new interacting modes emerge through a supercritical bifurcation. In this new regime, the system exhibits coherent subsystems of antiphase synchronized oscillators, which are stationary clusters following a spatiotemporal beating phenomena (ZZ regime). This regime is the result of a stronger coupling. We show from a phase mismatch model applied to each oscillator, that these phase coherent domains undergo a global phase instability meanwhile the interactions between oscillators become nonlocal. For each value of the control parameter we find out the time-varying topology (link matrix) from the contact interactions between oscillators.  The new characteristic spatiotemporal scales are extracted from the antiphase correlations at the time intervals defined by the link matrix. The interpretation of these experimental results contributes to widen the understanding of other complex systems exhibiting similar phase chaotic dynamics in 2D and 3D.

\bigskip


\noindent {\small{\it Note:} A version of this article has been accepted for publication on 12th January 2009 in the special issue "Modelling and Computation on Complex Networks" in the {\it Int. J. Bifurcation Chaos Appl. Sci. Eng.}} 

\smallskip

\end{abstract}

 \maketitle

\noindent {\bf 1. Introduction}

\noindent Complex systems, in nature, may undergo phase synchronization transitions driven by inertia, e.g. spatially extended multicellular systems of self-sustained oscillators in physics, chemistry, biology and neuroscience. The departure regime is a spatiotemporal chaos regime (STC) developed from a weak coupling between the units of the ensemble. However, under certain conditions, from this initial regime of STC, complex systems may display a simplified dynamics, which is described by fewer degrees of freedom, due to the effect of synchronization sustained by nonlinearities. For instance, in hydrodynamics, for highly developed turbulent flows, the interaction between vortices produces low-dimensional coherent structures like the large-scale vortices observed in von {K}\'arm\'an flow [de la Torre \& Burguete, 2007], or for weakly developed turbulence, far above the stationary rolls spirals arise in Rayleigh-B\'enard convection [Morris \textit{et~al.}, 1993],\par

 During the last years, the challenging field of complex networks has given rise to extensive reports [Boccaletti \textit{et~al.}, 2006] and renowned articles [Strogatz, 2001]. Experimental works in the field of complex networks with coupled nonlinear oscillators have been performed in: Josephson junction arrays [Fabiny \& Wiesenfeld, 1991], chemical oscillators in the Belousov-Zhabotinsky reaction [Fukuda \textit{et~al.}, 2005] and in the CO oxidation on platinum crystal [Falcke \& Engel, 1995], extended arrays of nickel electrodes in sulfuric acid [Kiss \textit{et~al.}, 2002], arrays of semiconductor lasers [Kozyreff \textit{et~al.}, 2000]. There are recent theoretical models performing extended multicellular systems with ring geometry [Abrams \& Strogatz, 2006; Rajesh \& Sinha, 2008].\par

Depending on the range of interactions between the oscillating units in a network, synchronization phenomena can be classified into: global synchronization (all-to-all), local synchronization (between nearest neighbors) and nonlocal synchronization (farther than the nearest neighbors). There are recent research works interested in time-dependent phenomena driven by nonlocal coupled oscillators [Tanaka \& Kuramoto, 2003; Shima \& Kuramoto, 2004]. A priori, under critical controlled conditions interactions between convective diffusively coupled cells are expected to be local.\par

The dynamics of complex system consisting of an ensemble of coupled oscillators, far enough from the first instability, can be described from their phases $\phi_i(t)$. This phase dynamics can shed light into the understanding of critical phenomena when a subsystem of oscillators bifurcates towards a new oscillatory state represented by an average phase $\Phi(t)$. This synchronized ensemble of oscillators defines a cluster.\par

First attempts on phase synchronization or self-entrainment theory were originally introduced by Winfree for biological communities [Winfree, 1967], and afterwards this theory was developed by Kuramoto [Kuramoto, 2003] in {\it the phase reduction theory of weakly coupled limit cycles} (KM) for an ensemble of N identical oscillators with global coupling. To characterize a phase synchronization transition, we look for an order parameter (or parameters) that quantifies the way in which the system evolves towards the new state.  
From the average synchronized phase of the system $\Phi$ (cluster phase synchronization), KM [Kuramoto, 2003] defines the order parameter $re^{i\Phi}$ for the ensemble of N oscillators with random phases $\phi_j$:
\begin{eqnarray}
\label{eq:Kuramotoorder} r e^{i\Phi}=\frac{1}{N}\sum_j e^{i\phi_j}
\end{eqnarray}

We report experimental results concerning a quasi-1D convective system whose behavior can be understood as the resulting dynamics of $N\approx 80$ convective cells ({\it oscillating thermal plumes}). These oscillators define a geometric network as they are placed on a spatially extended array. Coexisting with a stationary pattern (ST) of wavelength $k_s$, which imposes a discrete space symmetry, there are irregular and stationary clusters of time-dependent patterns where oscillators are preferably in antiphase (counterphase) with their nearest neighbors, the kind of limit-cycle oscillators. Similar patterns of clusters have been observed in the Belousov-Zhabotinsky reaction-diffusion system with global feedback [Vanag \textit{et~al.}, 2000].\par

The results reported here contribute to study time-varying topologies as we increase the driving force. We extract the topology from the dynamics, and we characterize the synchronization clustering process in the spatiotemporal beating regime (ZZ) from the antiphase matrix. Regarding topology, we find parallel contributions to the field of complex networks [Timme, 2006; Bialonski \& Lehnertz, 2006]. Other efforts towards the understanding of similar systems are developed in terms of dynamical weights from adaptive processes [Zhou \& Kurths, 2006], taking into account a varying coupling strength.\par

The aim of this paper is to understand the natural phenomenon of interacting oscillators in a 1D convective system. Inside the regime of spatiotemporal beats, the coupling between oscillators becomes nonlocal as the driving force is increased. This nonlocal coupling is responsible for a global transition deduced from the correlation times. For each control parameter value a time-dependent connectivity matrix is built. An order parameter is obtained from the cross-correlations between the critical phases ($\phi_i^c$) of N oscillators, and it is shown to be of the order of the spreading size of the new synchronized phase.\par

\bigskip

\noindent {\bf 2. The Convective Problem} \smallskip

Our convective layer of fluid is situated in a rectangular vessel ($L_x\times L_y$, $L_x\gg L_y$) opened to the atmosphere. The fluid used is transparent to visible light (silicone oil with a viscosity of 5 cSt). Quasi-1D-dynamics is caused by heating the fluid layer along a line which is placed underneath and at the center. This line is provided by placing a heater rail below the mirror that sustains the fluid layer. The temperature at the lateral walls in the larger direction is controlled at 20.0 $\pm$ 0.1 $^{\circ}$C, the same one as the room temperature. The array of convective cells is observed using the shadowgraphy technique which reveals the fluid dynamics inside the bulk of the layer on a screen. The control parameter is the vertical temperature difference, $\Delta T$, between the constant temperature along the heating line and the room temperature. Extended details on the experimental setup are found in a previous work [Miranda \& Burguete, 2008]. We have explored the range $\Delta T=[0-30]$ for depths of the fluid layer of $d=[2.5-5]$ mm. The corresponding stability diagram ($d$ vs $\Delta T$) has been already reported from the analysis of the spatiotemporal signals $S(x,t)$ [Miranda \& Burguete, 2008; Miranda \& Burguete, 2009]. At each asymptotic state, these signals are obtained by recording the shadowgraphy image along a line $x=[0,L_x]$ next to the center.\par

\begin{figure}[p]
\includegraphics[width =6.0in]{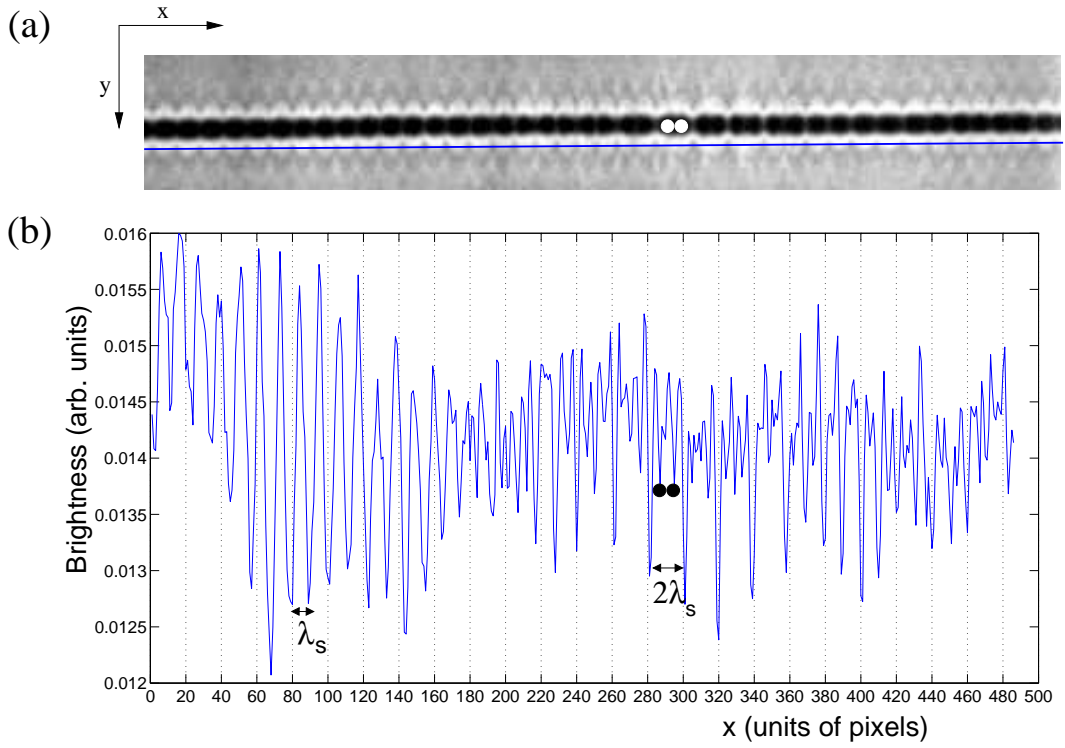}
\caption{(a) Photograph of the shadowgraphy image, dark spots at the center line correspond to the ascending hot plumes (hotspots). The continuous blue line represents the acquisition line recording the spatiotemporal diagrams $S(x,t)$. (b) Instantaneous brightness profile along the 1D-array (over the acquisition line). $\lambda_s$ is the wavelength of the stationary ST pattern. The bright spots (resp. dark spots) in the top (resp. bottom) figure correspond to the position of two adjacent oscillators.}\label{fig:figmirandanet1}
\end{figure}

In the shadowgraphy image [Fig.~\ref{fig:figmirandanet1}(a)], an oscillator will be associated to each one of these convective cells (or hotspots) along the heating line. The acquisition line to record the spatiotemporal signal $S(x,t)$ is represented in Fig.~\ref{fig:figmirandanet1}(a) by a blue line. This line is placed parallel to the aligned uprising hotspots in the center.
The correspondent brightness profile [Fig.~\ref{fig:figmirandanet1}(b)] allows to track the oscillators position in time. This is given by selecting the implied relative extrema from Fig.~\ref{fig:figmirandanet1}(b). We choose a characteristic spatial scale defined by the wavelength $2\lambda_s$ of the oscillating thermal cells in these instabilities.\par

\begin{figure}[p]
\includegraphics[width =6.0in]{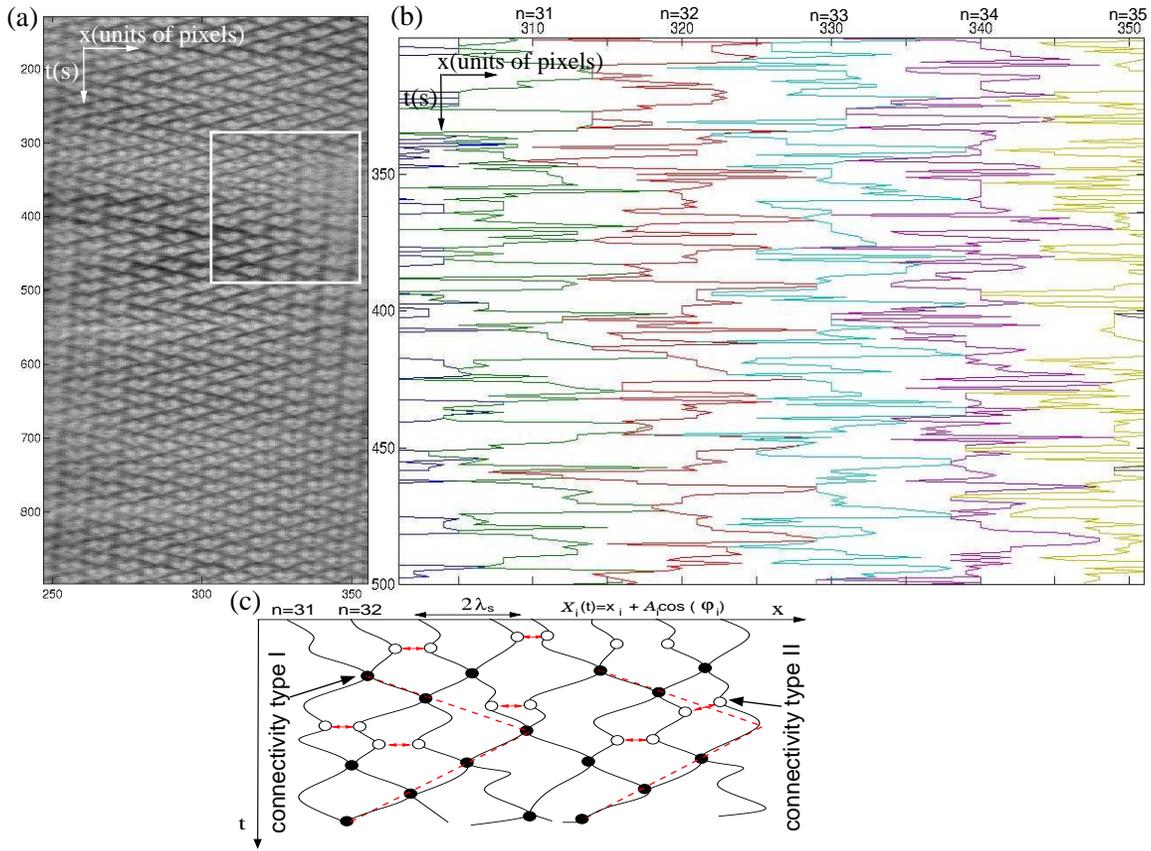}
\caption{(a) Shadowgraphy image at $\Delta T=$ 32 K; (b) trajectories of the oscillators tracked from the frame in (a); (c) sketch of the trajectories for a few oscillators separated at the characteristic spatial scale $2\lambda_s$ in the ZZ pattern. Black filled circles indicate the contact interaction loci (connectivity type I), white filled circles indicate the minimum distance between oscillators without collision (connectivity type II). Discontinuous lines are a guide to the eye showing the observed characteristic zig-zag (ZZ) pattern (temporal beats).  }\label{fig:figmirandanet2}
\end{figure}

In Fig.~\ref{fig:figmirandanet2}(a) we show a typical spatiotemporal diagram $S(x,t)$ above the threshold of the secondary bifurcation to a zig-zag regime, the ZZ pattern. This spatiotemporal diagram corresponds to the acquisition of global information about the interaction of N elements upon time. This global information, as it is explained later on in Sec. 3, can also be divided into N-oscillating units [see the corresponding tracking of N = 5 oscillators in Fig.~\ref{fig:figmirandanet2}(b)].
From each $S(x,t)$ diagram, bidimensional Fourier fast transform captures the essence of the global oscillations into a set of regional peaks. In the Fourier space, each peak corresponds to a mode with a well-defined ``spatiotemporal frequency band''. From the data analysis, for each fundamental mode we get $M(k,\omega
)$, where the wavenumber $k$ and the frequency $\omega$ are the maxima of each peak [Miranda \& Burguete, 2008; Miranda \& Burguete, 2009]. 

For a constant depth $d=$ 3 mm, as $\Delta T$ is increased beyond the ST regime, waiting for the asymptotic state to settle, the system goes through a ``mixed pattern'' in a STC regime. This STC regime is characterized by the presence of several irregular clusters. Oscillators belonging to this clusters perform an alternating pattern (ALT) [with wavenumber $k_{s}/2 \equiv k_{s/2}$ and with frequency $\omega_{alt}$] which coexists with the ST pattern in a mixed ST/ALT regime. From the coalescence of these various spreading phase subsystems the dynamics undergoes a supercritical bifurcation towards a new state of spatiotemporal beats (ZZ). This ZZ pattern is composed of two phase subsystems with high spatiotemporal coherence. It is characterized by a zig-zag geometry due to the splitting of the temporal frequency into ($\omega_{alt},\omega_{zz}$) with $\omega_{zz}\lesssim \omega_{alt}$. Also, there is a spatial frequency splitting into ($k_{s/2},k_{zz}$) with $k_{s/2}\lesssim k_{zz}$, which is responsible for the existence of two high coherent patterns. This ZZ regime calls for a phase description obtained from the signal of each oscillator. At the STC regime, weak coupling can sustain irregular clusters meanwhile, in the ZZ regime, a stronger coupling locks the phase into stationary clusters. From the diagrams $S(x,t)$, we have shown in a previous work [Miranda \& Burguete, 2009] the evolution of the fundamental modes in the mixed ST/ALT and the ZZ regimes, and also that the oscillating frequencies increase linearly with the control parameter.
For thicker layers the oscillators are able to synchronize moving towards one privileged direction, a travelling wave regime (TW) [Miranda \& Burguete, 2008].\par

Phase mismatches between 50 oscillators, calculated using a spatiotemporal cross-correlation signal processing, provide the framework to understand the spatiotemporal coherence of the beating regime from a nonlocal coupling. New dynamical conditions like stronger nonlinear coupling set a longer interaction range, further than first neighbors (three interacting oscillators). An increased number of interacting oscillators is the result of strongly interacting convective cells, we might think of a new structured flow that encloses several long-ranged interacting cells.\par

On the other hand, connection topologies are not fixed because the topology of the network is defined by time-varying links between the oscillators. The loci of the links are determined from the collapsing interactions between the nearest neighboring oscillators (link matrix). In these contact interactions two convective cells are colliding. \par

Also, we measure the synchronization range (number of synchronized oscillators) given by the correlated synchronized phases $\phi_i$. These phases are obtained demodulating the raw phase $\Psi_i$ of each oscillator nearby the selected synchronization frequency ($\omega_i=\partial_t\phi_i$). These raw phases $\Psi_i$ are functions of the several unstable or critical temporal modes.\par

Hence, in our 1D-array, a ``critical phase'' ($\phi_i^c$) emerges at the threshold of the synchronization transition for each oscillator. This critical phase $\phi_i^c$ slightly shifts from the cluster phase $\Phi$. Under these conditions, the developed analysis techniques reported here are somehow inspired by the Eq.(~\ref{eq:Kuramotoorder}) (KM). The key is to build a minimum cross-correlation matrix (antiphase matrix) and the corresponding matrix of time lags at minimum correlation ($\equiv$ maximum antiphase correlations) between each pair of oscillators.\par

\bigskip

\noindent {\bf 3. Spatiotemporal Phase Synchronization Transition Scenario} \smallskip

\noindent In our ensemble of oscillators, the phase chaotic regime extends over the STC and ZZ regimes. Firstly, short-range interactions (due to diffusive coupling) are able to develop a regime of STC. In this regime, spatiotemporal correlations in the stationary pattern decay due to the increasing presence of irregular clusters. This chaotic dynamics corresponds to an inhomogeneous growth of the ALT pattern until a critical value [Miranda \& Burguete, 2009]. Secondly, a phase synchronization regime is characterized by the presence of beats, ZZ (see the sketch for the temporal beats in Fig.~\ref{fig:figmirandanet2}(c)). These beats are due to a small shift in the spatial and temporal frequencies. From now on, we will focus on the latter synchronization transition. Experimental results follow this ZZ regime for an ascending sequence of 5 K by steps of 0.5 K. Spatial desynchronization appears for higher control parameter values with the presence of phase singularities. Thus, a temporal synchronization only remains.\par

When the original signal along the entire ensemble $S(x,t)$ is decoded for $N=$ 50 oscillators [e.g. for a few oscillators from the N-ensemble see Fig.~\ref{fig:figmirandanet2}(b)] we obtain an individual signal for each $i$-oscillator: $\mathcal X_i(t)=x_i+\mathcal A_i \cos[\Psi_i(\phi_i^{alt},\phi_i^{zz},\phi_i^\circ)]$, where $\Psi_i$ is the raw phase containing all the involved critical phases $\phi_i^{c}$ (for the ALT and ZZ characteristic phases), and $\phi_i^\circ$ is a constant phase shift. In this way, we are able to understand how do the oscillators interact in space and time. 

\begin{figure}[p]
\includegraphics[width =6.0in]{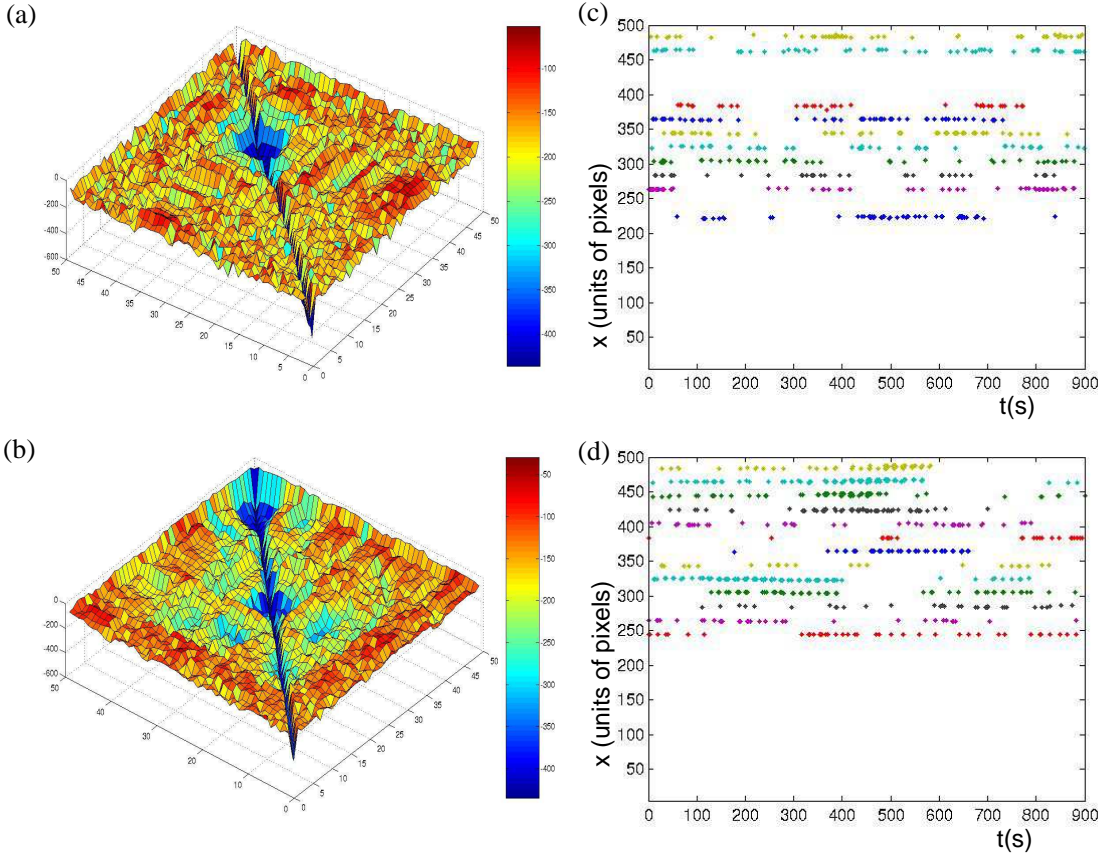}
\caption{Surface views of the antiphase correlation matrices $A_{ij}(\omega_{zz})$ at the critical frequency mode $\omega_{zz}$ for two phase synchronization states at (a) $\Delta T=\Delta T_c=31$ K and (b) $\Delta T=34$ K. The corresponding link matrices $L_{ij}$ generated by contact interactions between oscillators are shown in pictures (c) for $\Delta T=\Delta T_c=31$ K and (d) $\Delta T=34$.}\label{fig:figmirandanet3}
\end{figure}

We demodulate $\Psi_i$ around the emerging unstable mode $\omega_i^{zz}=\partial_t\phi_i^{zz}$, the new phase associated to each oscillator is simply given by $\phi_i(\equiv \phi_i^{zz})$. The cross-correlation phase vector is given by:
\begin{eqnarray}
\label{eq:corr} \mathcal C_{ij}(\omega_{zz},\tau_l)=\langle \cos(\phi_i^{zz})\cos(\phi_j^{zz})\rangle
\end{eqnarray}
where $\tau_l$ are the time lags, e.g. if we take a couple of oscillators $(i,j)$ in antiphase, this behavior belongs to the characteristic alternating pattern ALT (two counteroscillating modes with $\pm \omega_{alt}$), then $\tau_l$ is given by the half of a period $T$ ($T/2$ is the time that takes the $i$-oscillator $\mathcal X_i(t)$ to catch up the j-oscillator $\mathcal X_j(t+T/2)$). 
Because the great majority of oscillators are in antiphase, the cross-correlation vector $\mathcal C_{ij}$ happens to show the strongest correlation between neighboring oscillators in the minima of the cross-correlation vector (antiphase conditions). This fact naturally comes from the type of attractive-repulsive interaction between oscillating units. In order to extract information about the phase correlations along the array we study this antiphase matrix [see Fig.~\ref{fig:figmirandanet3}(a,b)] defined as:

\begin{eqnarray}
\label{eq:ac} A_{ij}(\omega_{zz})={\displaystyle \min_{\tau_l}}\;\mathcal C_{ij}(\omega_{zz},\tau_l)
\end{eqnarray}

This is a $N\times N$ matrix whose elements are given as the minimum of $\mathcal C_{ij}$ (maximum antiphase correlations) at their corresponding time shifts $\tau_l$.
Conclusively, we may consider that this matrix is actually supplying the ``degree of phase shifts'' between each oscillator regarding the cluster synchronization phase: $\left|\phi_i-\Phi\right|$. In this way, the cluster phase $\Phi$ fulfills the following definition $\partial_t\Phi(t)=\omega_{zz} \equiv \sum_{i=1}^n \omega_i^{zz}/n$, where $\omega_{zz}$ is the average critical mode of $n\le N$ synchronized oscillators. As we show in Fig.~\ref{fig:figmirandanet3}(a,b) the surface representation of the matrix $A_{ij}(\omega_{zz})$ contains the coupling nature between all-to-all oscillators of a 1D array. In these figures the depressed regions (the deeper ones) correspond to the higher synchronized oscillators defining the boundaries that surround clusters. Hence as we increase the control parameter a new phase synchronization domain emerges which correspond to the spatial beating phenomena: two stationary clusters [Fig.~\ref{fig:figmirandanet3}(b)].\par

Along the ascending sequence in $\Delta T$, from the analysis of $\tau_l$ applied to an oscillator belonging to the cluster ($n=$ 31), it has not been possible to establish any dependence on $\Delta T$. From this point on, our research follows the information given the antiphase matrix, $A_{ij}(\omega_{zz})$ whose values inside a cluster keep sufficiently small, below a fixed correlation value $A_{ij}(\omega_{zz})=\epsilon$ (for this sequence $\epsilon=- 200$ [Fig.~\ref{fig:figmirandanet3}(a,b)]). Therefore, $\epsilon$ is kept constant in order to guarantee the phase locking between neighboring oscillators for the whole oscillators array and for a whole sequence of measurements. When the first stationary cluster emerges from $A_{ij}(\omega_{zz})$ the critical control parameter is $\Delta T_c=31$ K [Fig.~\ref{fig:figmirandanet3}(a)].\par
We define the order parameter $\mathcal Z(\omega_{zz})$ to quantify the dynamics of the complex network as we increase the control parameter. Firstly we define for each $i$-oscillator $\mathcal Z_i(\omega_{zz})$ (as we move from the left to the right): 

\begin{eqnarray}
\label{eq:defZi} \mathcal Z_i(\omega_{zz})=\sum_{j=i+1}^{\eta \le N}  f_{ij}, \; \mbox{where}  \; f_{ij}= \left\{ \begin{array}{cc} 1 \hspace{5mm} A_{ij}(\omega_{zz})< \epsilon \\  0 \hspace{5mm} A_{ij}(\omega_{zz})> \epsilon\end{array}      \right.
\end{eqnarray}

\begin{figure}[p]
\includegraphics[width =4.0in]{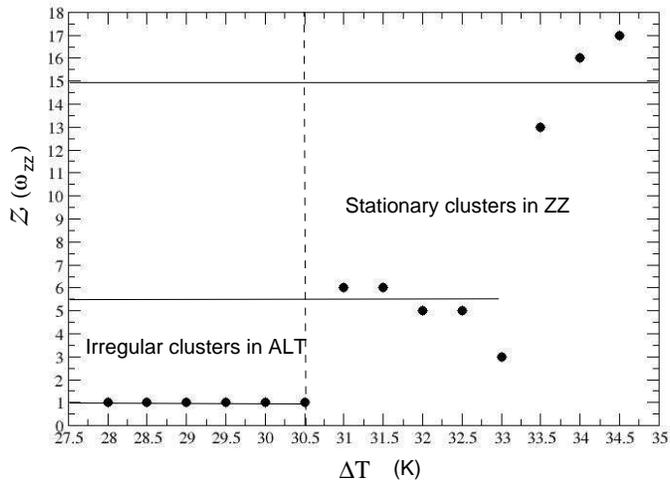}
\caption{Order parameter values $\mathcal Z(\omega_{zz})$ for the ascending sequence. The vertical dashed line split two different regimes. Horizontal solid lines show averaged values of the number of oscillators involved in clusters.} \label{fig:figmirandanet4}
\end{figure}

The result given by $\mathcal Z_i(\omega_{zz})=\eta$ means that for a given $i$-oscillator, the following number $\eta \le N$ of oscillators are highly correlated to it. From this analysis we are able to determine the maximum number of synchronized oscillators in a cluster at the asymptotic state determined by $\Delta T$. Thus, the order parameter is obtained from:

\begin{eqnarray}
\label{eq:defZ} \mathcal Z(\omega_{zz})={\displaystyle \max_i} \left(\mathcal Z_i\left(\omega_{zz}\right)\right)
\end{eqnarray}

for an $i$-oscillator belonging to the cluster. In Fig.~\ref{fig:figmirandanet4} we find out, far from the threshold ($\Delta T>33$ K) and for the spatial coherent subsystem (stationary cluster in ZZ), that 15-17 oscillators are coupled, but for the same pattern and lower control parameter values approximately 5 oscillators are coupled. \par
\begin{figure}[p]
\includegraphics[width =4.0in]{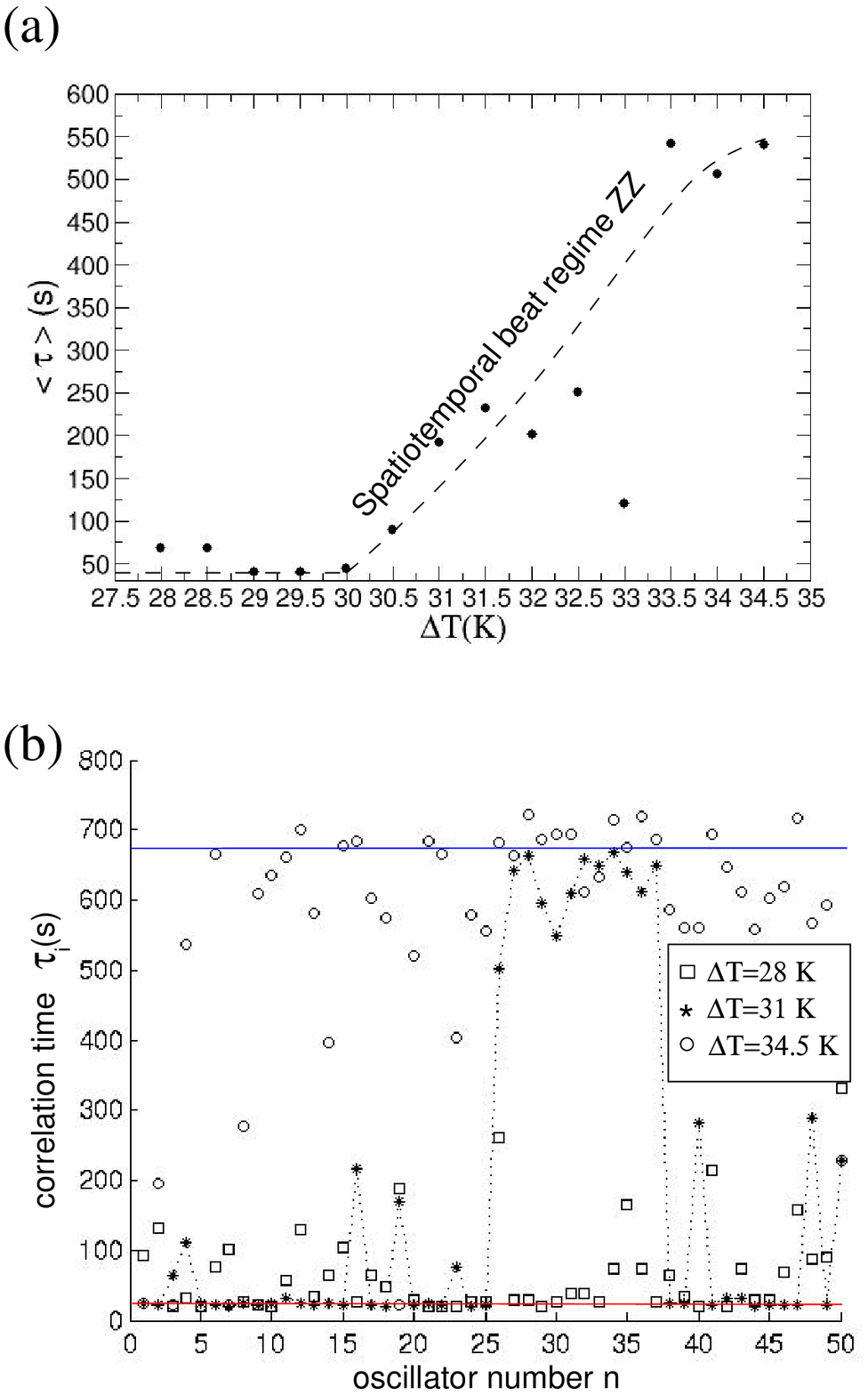}
\caption{(a) Average correlation time $\langle\tau\rangle$ vs $\Delta T$, discontinuous line is a guide to the eye; (b) correlation time ($\tau_i$) for each oscillator at $\Delta T=$28, 31, 34.5 K. Top and bottom continuous lines represent the maximum and minimum time correlations, dotted lines for $\Delta T=$31 K are a guide to the eye.} \label{fig:figmirandanet5}
\end{figure}

The outstanding presence of the spatial coherent domains (spatial beats) as the system moves forward a global synchronization instability is shown from the average correlation time $\langle\tau\rangle$ [Fig.~\ref{fig:figmirandanet5}(a)]. The assigned average value $\langle\tau\rangle$ for each control parameter value is obtained from the correlation times of the whole array ${\tau_i}$. These are measured from the maximum time interval at which $|\mathcal C_{ij}(\omega_{zz},\tau_l)|$ decays to $|\mathcal C_{ij}(\omega_{zz},\tau_l)| e^{-1}$ for $\tau_l=0$. In Fig.~\ref{fig:figmirandanet5}(b) we show the distribution of $\tau_i$ for three representative control parameter values. Far from the threshold (at $\Delta T=34.5$ K), $\tau_i$ reaches the top time recording value for almost all the oscillators, in contrast to the values obtained below the threshold (at $\Delta T=28$ K). Clearly, at the threshold ($\Delta T_c$) the oscillators belonging to the cluster show the longest correlation times [Fig.~\ref{fig:figmirandanet5}(b)].\par

\begin{figure}[p]
\includegraphics[width =4.0in]{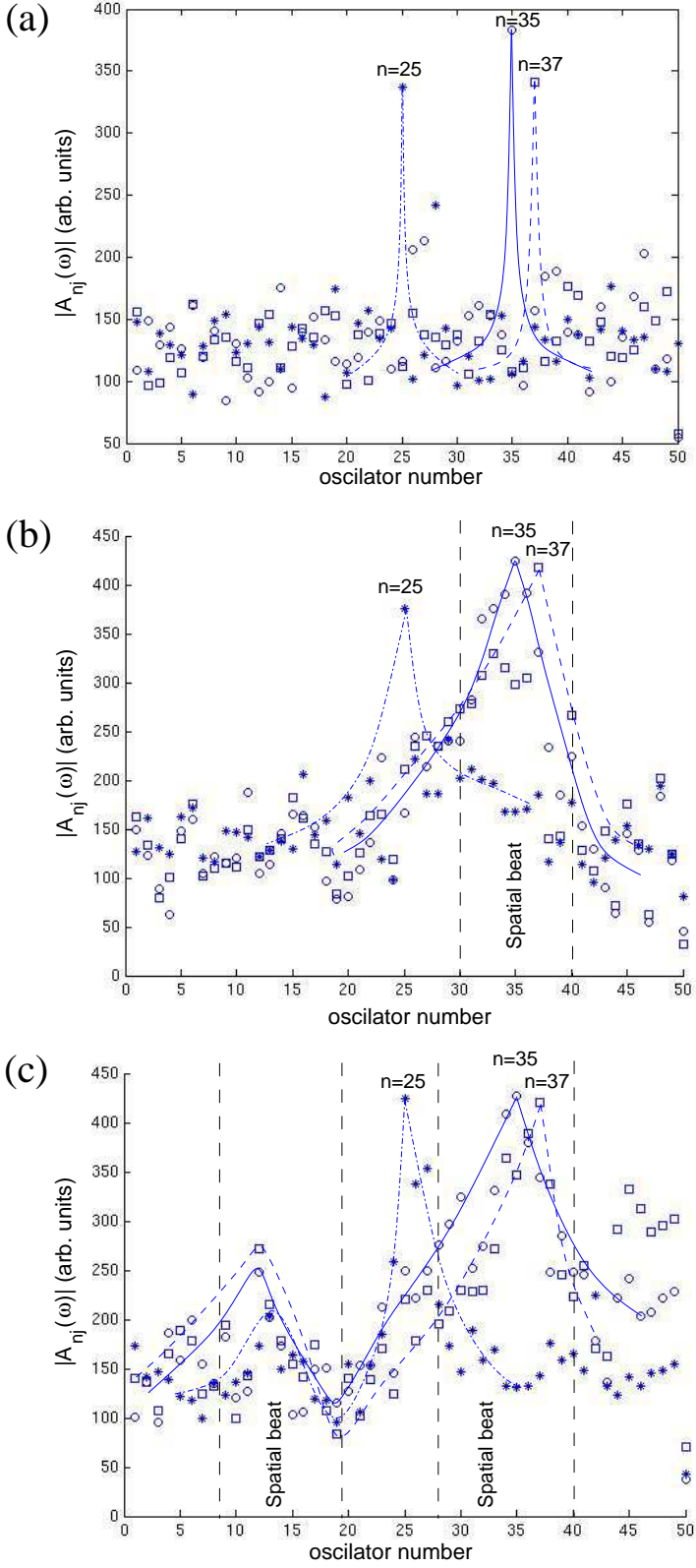}
\caption{Antiphase cross-correlations $A_{nj}(\omega_{zz})$ ($j=1,\ldots,N$) for oscillators $n=$ 25, 35, 37 at: (a) $\Delta T=$ 28 K; (b) $\Delta T=$ 31 K and (c) $\Delta T=$ 34.5 K . Continuous and dashed lines are a guide to the eyes. Vertical dashed lines define the cluster domain.} \label{fig:figmirandanet6}
\end{figure}

But there is still one missing piece in this network puzzle, when we ask about what is going on with the interaction range in the overall ascending sequence towards this global synchronization instability. Regarding a nonlinear hydrodynamics point of view, we might think of a new dynamics of the convective cells in the phase chaotic regime, a kind of package $\Delta n$ consisting of more than three interacting oscillators. This wider ``enveloping convective cell '' could not be the effect of a merely diffusive local coupling as for the ST regime. The ZZ regime might be sustained by a stronger nonlinear coupling between oscillators. If we choose three oscillators from the array [see Fig.~\ref{fig:figmirandanet6}(a-c)]: $n=$ 25 (this oscillator belongs to the front of the right hand side cluster only far from the threshold [Fig.~\ref{fig:figmirandanet6}(c)]) and $n=$ 35,37 (these ones belong to the cluster on the right hand side [Fig.~\ref{fig:figmirandanet6}(b,c)]), then representing $|A_{nj}(\omega_{zz})|$ for a fixed $i=n$ and $j=1,\ldots,N$, from Fig.~\ref{fig:figmirandanet6}(a-c), we find out higher phase correlation values, and therefore stronger coupling, the further the control parameter is from the threshold. Besides, the decay tendency of the antiphase cross-correlations between the selected oscillators and the nearest neighbors is smoother far from the threshold [Fig.~\ref{fig:figmirandanet6}(c)], considering the abrupt decay of $|A_{nj}(\omega_{zz})|$ between nearest neighbors in Fig.~\ref{fig:figmirandanet6}(a). We observe how this strong coupling is already outstanding at the threshold [Fig.~\ref{fig:figmirandanet6}(b)]. These last results might allow us to determine an effective nonlocal interaction range $\Delta n$ from the cross-correlations. Nevertheless, the topology (link matrix) will be involved in these results, as it will be shown in the next section, from the colliding positions between oscillators.\par

\bigskip

\noindent {\bf 4. Time-Varying Topology } \smallskip
\label{sec:topology}

\noindent Our objective is to decode some topological aspects such as geometric loci of the oscillators and the degree of connectivity from the dynamics of our 1D coupled network. It is interesting to consider the degree of attractive interaction between oscillators, regarding this fact we could define two kinds of connectivities [sketch in Fig.~\ref{fig:figmirandanet2}(c)]: type I for an overlapped pair of oscillators, a collision between two adjacent hotspots has been produced; and type II when the distance between two neighboring oscillators reaches a minimum but without collision. In the following, connectivity always refers to type I.\par

In the synchronized regime ZZ, we build the link matrix $L_{ij}$ by locating in the tracking image [e.g. Fig.~\ref{fig:figmirandanet2}(b)]  the contact loci between two adjacent antiphase oscillators that satisfy: $\mathcal X_{i-1}(t)=\mathcal X_i(t)$ [see sketch in Fig.~\ref{fig:figmirandanet2}(c)]. A link point is a collision position, this means that a couple of oscillators overlap at the same time that the linking is affecting at least 4 oscillators.  After filtering contact links provided only by oscillators with frequency $\omega_{zz}$, $L_{ij}$ is nonzero only at these contact loci [e.g. Fig.~\ref{fig:figmirandanet3}(c,d)]. The analysis of  $L_{ij}$ shows that as the system is getting far from the threshold the number of links increases. Moreover, it allows to study how the dynamics of the rewiring connections, for different control parameter values $\Delta T$, plays a role in the coupling interaction between oscillators.\par
\begin{figure}[p]
\includegraphics[width =6.0in]{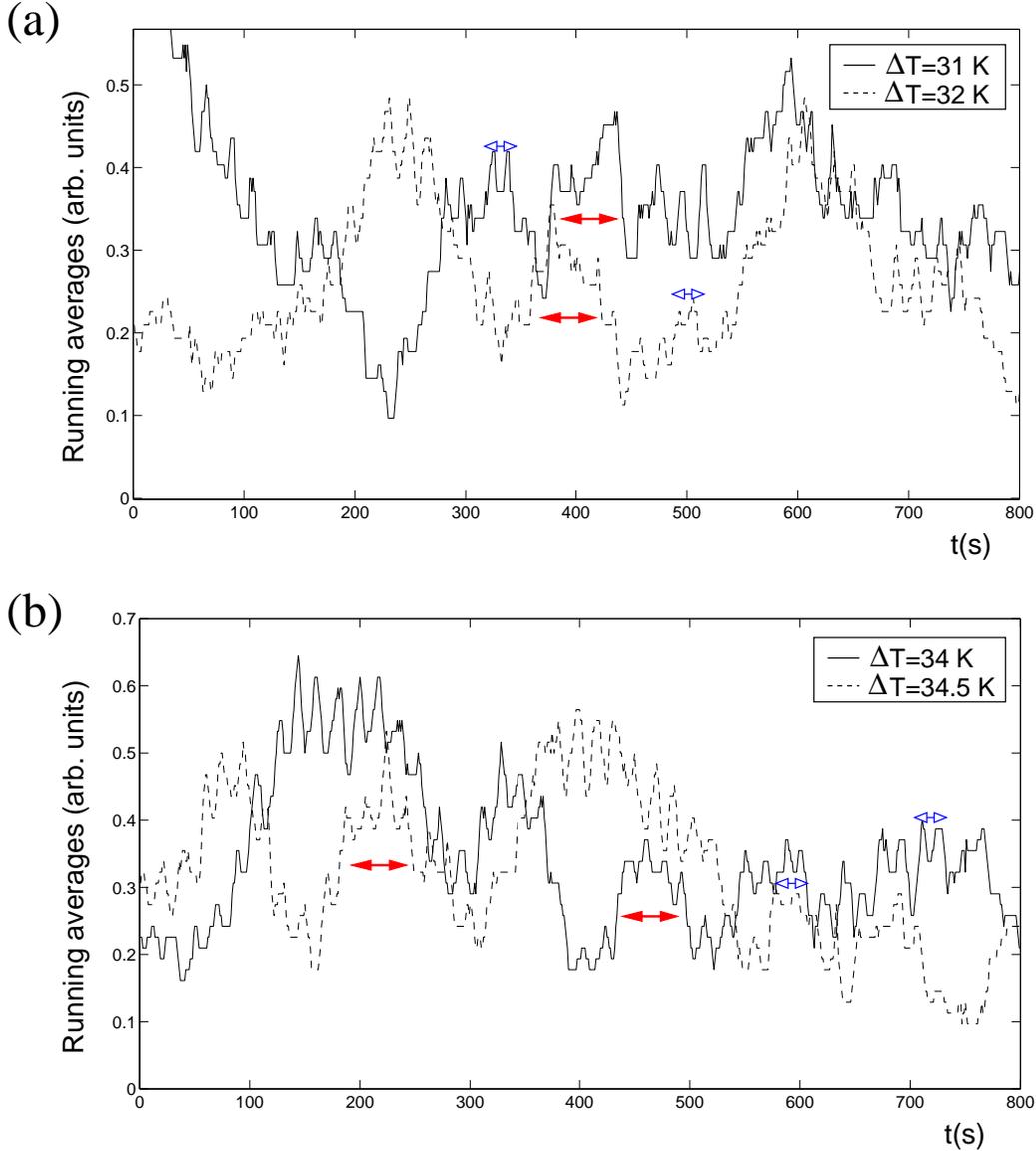}
\caption{(a) Running averages over 62 seconds of the link matrix $L_{ij}$ at $\Delta T=$31,32 K; (b) running averages over 62 seconds of the link matrix $L_{ij}$ at $\Delta T=$34, 34.5 K. Red and blue doubled arrows show the temporal beat period (envelope) and the internal wave (carrier) period respectively.} \label{fig:figmirandanet7}
\end{figure}
\begin{figure}[p]
\includegraphics[width =7.0in]{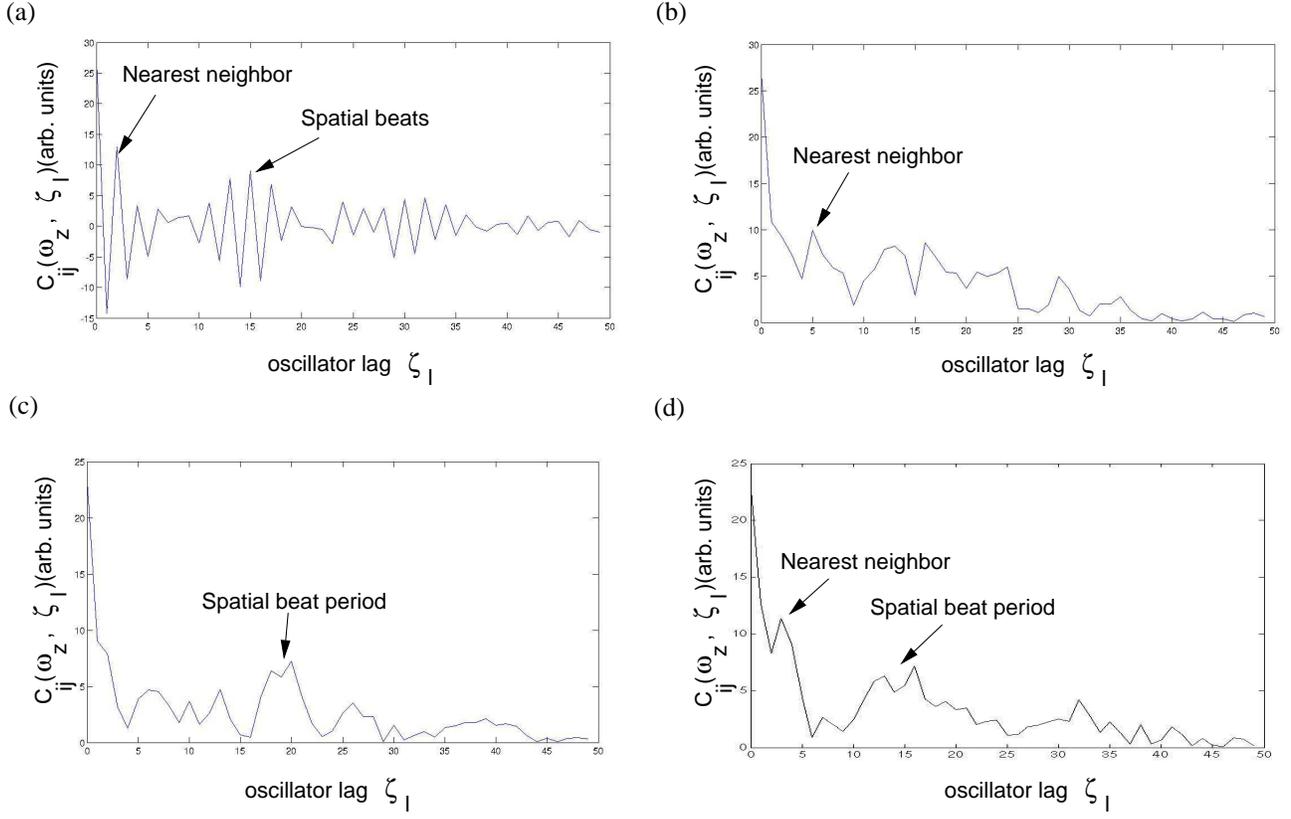}
\caption{(a) Cross-correlation vector at time of maximum linking, $\mathcal C_{ij}(\omega_{zz},\chi_l)|_{t=404s}$ ($t=404$ s time of maximum linking determined by the maxima of link matrix $L_{ij}$) for $\Delta T=34.5$ K . Module of the spatial correlations: (b) at minima $|\mathcal C_{ij}(\omega_{zz},\chi_l)|_{t=100s}$ for $\Delta T=33.5$ K; (c) at maxima $|\mathcal C_{ij}(\omega_{zz},\chi_l)|_{t=797s}$ for $\Delta T=34.5$ K; (d) at maxima $|\mathcal C_{ij}(\omega_{zz},\chi_l)|_{t=61s}$ for $\Delta T=33.5$ K .} \label{fig:figmirandanet8}
\end{figure}

The link matrix $L_{ij}$ provides quantitative information about the evolution of the number of links in time. This information is simplified by taking running averages over a time interval [Fig.~\ref{fig:figmirandanet7}]. We choose 62 s as the time to average, which approximately corresponds to half of the periodicity of the temporal beating. From $L_{ij}$ along the ascending sequence, we recover the characteristic time periods from these graphics [Fig.~\ref{fig:figmirandanet7}]. The envelope of the temporal beat and the internal wave (carrier) periods are approximately 120 s and 15 s respectively [Fig.~\ref{fig:figmirandanet7}(a,b)]. At this point, we look for an homologous cross-correlation phase behavior at times belonging to the minima ( $t_m$ with a lower degree of links), and maxima ($t_M$ with a higher degree of links) of the running averages over $L_{ij}$. In order to study this effect we obtain the cross-correlation phase vector for a fixed time:
\begin{eqnarray}
\label{eq:spatialcorr} \mathcal C_{ij}(\omega_{zz},\chi_l)=\langle \cos(\phi_i(t_{m,M})),\cos(\phi_j(t_{m,M}))\rangle
\end{eqnarray}
where $\chi_l$ is the oscillators lag $\chi_l=1,\ldots, N$. When the number of links is zero the ``spatial'' cross-correlation phase vector is chaotic, meanwhile when the number of links is nonzero the typical cross-correlation vector is similar to Fig.~\ref{fig:figmirandanet8}(a). From the absolute value $|\mathcal C_{ij}(\omega_{zz},\chi_l)|_{t_{m}}$ we identify a correlation length of 4-5 oscillators [Fig.~\ref{fig:figmirandanet8}(b)] with low peaks at the cluster size (13-17 oscillators belonging to the clusters). Otherwise, from the absolute value $|\mathcal C_{ij}(\omega_{zz},\chi_l)|_{t_{M}}$ we identify the maximum correlation ``length'' at the cluster size of approximately 15-18 oscillators [Fig.~\ref{fig:figmirandanet8}(c)]. But certain cross-correlation behavior, like the one showed in Fig.~\ref{fig:figmirandanet8}(d), still shows the maximum coupling strength around 4 oscillators. Thus, far from the STC regime, in the ZZ phase synchronization transition, we are allowed to consider that the dynamics affects the underlying topology of the network.\par
From these results, it follows that nonlocal coupling existing along the ZZ regime could be possible. This coupling might go beyond first neighbors ($\Delta n >3$), differing from the kind of coupling that it was already shown below the threshold in Fig.~\ref{fig:figmirandanet6}(a) for $\Delta n=$ 3. In the ZZ regime, because the interaction range is wider than a purely diffusive one we might certainly expect roughly $\Delta n= 4-5$ oscillators. \par

\bigskip

\noindent {\bf 5. Discussion and Conclusions} \smallskip

\noindent We have reported the study of a synchronization transition beyond spatiotemporal chaos in a hydrodynamic system. This consists of an ensemble of $N\approx 50$ identical oscillators (the size of each oscillator is approximately of 6 mm in a real array 80 oscillators), developed in a fluid layer under 1D thermal convection [Miranda \& Burguete, 2009]. These coupled $i$-oscillators $(i=1,\ldots,N)$ whose signals are given by: $\mathcal X_i(t)=x_i+\mathcal A_i \cos[\Psi_i(\phi_i^{alt},\phi_i^{zz},\phi_i^\circ)]$, are placed in the array at their respective ``zero positions'' $(x_i)$, which settle two very close wavenumbers: $k_{s/2}$ and  $k_{zz}$; and have two close critical frequencies: $\omega_{alt}$ and $\omega_{zz}=\partial_t\Phi$, respectively. We have described the synchronization process in terms of the individual phases $\phi_i$ regarding the cluster phase $\Phi$, that is, the analysis of the antiphase matrix $A_{ij}(\omega_{zz})$. The surface of $A_{ij}(\omega_{zz})$ defines the stationary cluster domains for the critical mode $\omega_{zz}$, in addition it shows no significant correlation for $\omega_{alt}$. The number of synchronized oscillators is defined by $\mathcal Z_i(\omega_{zz})=\eta < N$ for $i=1,\ldots,N$ and agrees with the assigned high time correlation values $\tau_i$.\par

We might notice that from the time lag analysis between each pair of oscillators $(i,j)$, defined as $\tau_{l,i}\equiv \tau_{l,j}\left(mod\;T_{zz}\right)$ (module $T_{zz}=2\pi/\omega_{zz}$), is more difficult to recover the number of synchronized oscillators ($\mathcal Z_i(\omega_{zz})$) than from the analysis of the antiphase matrix.  Furthermore, the averaged value of time lags (in the ascending sequence) between an oscillator belonging to the cluster regarding the remaining oscillators, show no connection with the increasing time correlations.\par

The values of $\mathcal Z_i(\omega_{zz})$ for an $i$-oscillator belonging to the cluster on the left are less well-defined than the ones from the cluster on the right, and the corresponding amplitude inside the cluster happens to be ``turbulent'' for the critical mode $\omega_{zz}$. It should be taken into account that the order parameter $\mathcal Z(\omega_{zz})$ depends on the phase distribution along the 1D-array. From Fig.~\ref{fig:figmirandanet4} we show that the number of coupled oscillators belonging to the cluster increases from 5 to 17 for higher values of the control parameter. Although the frequency is a monotonically increasing function with temperature $\omega_{zz}(T)$ [Miranda \& Burguete, 2009], $\mathcal Z(\omega_{zz})$ is not a continuous function of $\Delta T$ . At $\Delta T=$ 33 K, the order parameter and average time correlation correspond to a desynchronized pattern.\par

Results on Fig.~\ref{fig:figmirandanet6}(a-c), obtained from the antiphase matrix, shows the type of the coupling interaction between the oscillators at the boundary and the oscillators belonging to the synchronized domain. In Fig.~\ref{fig:figmirandanet6}(b), it is shown that for the most distant oscillator ($n=$ 25) from the boundary of the synchronized ZZ pattern, the phase correlation decays ``more quickly'', as it is expected from a diffusive coupling of the type $A_{nj}(\omega_{zz})\sim e^{-\beta x}$ (short interaction range). Meanwhile for the remaining oscillators  belonging to the cluster, the antiphase correlation has a smoother decay. This fact should be interpreted as the consequence of a longer interaction range in an ``envelope roll'' of $\Delta n= 4-5$ interacting oscillators.\par

In our spatially extended array, we have shown the dependence of the network topology on the dynamics [Fig.~\ref{fig:figmirandanet7}]. The increasing number of contact interactions provide a sensitive dynamics which is revealed by the peaks in the spatial correlations between oscillators  at $|\mathcal C_{ij}(\omega_{zz},\chi_l)|_{t_{m,M}}$ [Fig.~\ref{fig:figmirandanet8}(a-d)], also we get the size of the clusters.\par

We have shown that the wiring of connections from the link matrix $L_{ij}$, is not fixed, neither in space nor in time. From the phase description theory (KM), a suitable model for an ensemble of identical $i$-oscillators, in the framework of nonlocal two-way coupling for attractive-repulsive interactions, could be $(i=1,\ldots,N)$:

\begin{eqnarray}
\label{eq:phase1} \dot{\phi_{i}}=F_{i}(\phi_i^{zz},\phi_i^{alt},\ldots,\phi_i^\circ)+\sum_{j=1}^N H_{ij}(t)\cdot \Gamma_{ij}(\phi_i-\phi_j)
\end{eqnarray}

Each $i$-oscillator has its own dynamics represented by $F_{i}$ as a function of the critical phases. The time-dependent connectivity function is the adjacency matrix $H_{ij}(t)=M_{ij}\cdot G_{ij}(t)$. From $\Delta n$ (given by the analysis of the running averages of the link matrix $L_{ij}$) we can build an adjacency matrix at zero time $M_{ij}(0)$ such that $M_{ij}=1$ if $|i-j|\le \Delta n$, and $M_{ij}=0$ otherwise.  The time-dependent connectivity, extracted from $L_{ij}$, might be mapped into $G_{ij}(t)$. The coupling function between oscillators is $\Gamma_{ij}(\phi_i-\phi_j)$, for this type of model equation Eq.(\ref{eq:phase1}) comes from a slow perturbation (e.g. with $\Gamma_{ij}(\phi_i-\phi_j)\simeq \sin(\phi_i-\phi_j)$ [Kuramoto, 1975]), the real coupling function in the ZZ regime might have a more complex behavior according to a varying coupling strength that depends on the control parameter $\Gamma_{ij}(\phi_i-\phi_j, \Delta T)$.\par

For a supercritical bifurcation, a phase description might not necessarily hold for any regime far from the first instability, because the possible coupling between the amplitude and the phase should play a role. Regarding this fact, synchronous behavior in the array is related to the existence of an unstable traveling mode $\omega_{zz}$ with enough amplitude to start synchronization (keeping strong values in $|A_{ij}(\omega_{zz})|$) sustained by nonlinearities. When this critical mode $\omega_{zz}$ is missing the rewiring of connections is only due to the mode $\omega_{alt}$. This mode represents weak coupling for counteroscillating modes, and moreover the corresponding interaction range still represents the kind of diffusive transport (of heat and viscous momentum) that one may expect from a local coupling [Fig.~\ref{fig:figmirandanet6}(a)]. In conclusion, the reported results for an spatially extended system show how the amplitude of the unstable mode plays a role in the phase synchronization process (e.g. according to numerical works  [Vadivasova \textit{et~al.}, 2001]). Besides, nonlocal coupling could explain similar transitions in nature and experiments, as it has been reported in recent theoretical works [Abrams \& Strogatz, 2006]. This synchronization phenomena is a new challenge that might need to be considered in the theory of cluster formation.\par

\bigskip

\noindent {\bf Acknowledgments} \smallskip

\noindent We are grateful to W. Gonz\'alez--Vi\~nas for fruitful discussions and comments. We are also indebted to H. Mancini for encouraging this research. This work has been partly supported by the Spanish Contract No. BFM2002-02011 and No. FIS2007-66004, and by PIUNA (University of Navarra, Spain). M.A. Miranda acknowledges financial support from the ``Asociaci\'on de amigos de la Universidad de Navarra''.

\bigskip

\newpage

\noindent {\bf References} \smallskip

\noindent Abrams, D.M. \& Strogatz, S.H. [2006]  "Chimera states in a ring of
  nonloclly coupled oscillators,"
\newblock \textit{Int. J. {B}ifurcation and {C}haos} \textbf{16}(1), 21--37.

\noindent Bialonski, S. \& Lehnertz, K. [2006]  "Identifying phase synchronization
  clusters in spatially extended dynamical systems,"
\newblock \textit{Phys. Rev. E} \textbf{74}, 051909.

\noindent Boccaletti, S., Moreno, Y., Chavez, M. \& Hwang, D.-U. [2006]  "Complex networks:
  structure and dynamics,"
\newblock \textit{Physics Reports} \textbf{424}, 175--308.

\noindent de la Torre, A., Burguete, J. [2007]  " Slow dynamics in a turbulent von {K}\'arm\'an swirling flow,"
\newblock \textit{Phys. Rev. Lett.} \textbf{99}, 054101.

\noindent
Fabiny, L. \& Wiesenfeld, K. [1991]  "Clustering behaviour of oscillator
  arrays,"
\newblock \textit{Phys. Rev. A} \textbf{43}, 2640--2648.

\noindent
Falcke, M. \& Engel, H. [1995]  "Cluster formation, standing waves, and
  stripe patterns in oscillatory active media with local and global coupling,"
\newblock \textit{Phys. Rev. E} \textbf{52}, 763--771.

\noindent
Fukuda, H., Morimura, H. \& Kai, S. [2005]  "Global synchronization
  in two-dimensional lattices of discrete {B}elusov-{Z}habotinsky oscillators,"
\newblock \textit{Physica D} \textbf{205}, 80--86.

\noindent
Kiss, I.Z., Zhai, Y. \& Hudson, J.L. [2002]  "Emerging coherence in a
  population of chemical oscillators,"
\newblock \textit{Science} \textbf{296}, 1676.

\noindent
Kozyreff, G., Vladimirov, A.G. \& Mandel, P. [2000]  "Global coupling
  with time delay in an array of semiconductor lasers,"
\newblock \textit{Phys. Rev. Lett.} \textbf{85}, 3809--3812.

\noindent
Kuramoto, Y. [1975] in \textit{International Symposium on Mathematical Problems in Theoretical Physics}, ed. Araki, H., Lecture Notes in Physics Vol. \textbf{30} (Springer, New York), p.420.

\noindent
Kuramoto, Y. [2003] \textit{Chemical oscillations, waves and turbulence} (Dover).

\noindent
Miranda, M. A. \& Burguete, J. [2008]  "Subcritical
  instabilities in a convective fluid layer under a quasi-one-dimensional
  heating,"
\newblock \textit{Phys. Rev. E} \textbf{78}, 046305.

\noindent
Miranda, M. A. \& Burguete, J. [2009]  "Experimentally observed route to spatiotemporal chaos in an extended one-dimensional array of convective oscillators,"
\newblock \textit{Phys. Rev. E} \textbf{79}, 046201.

\noindent
Morris, S. W., Bodenschatz, E., Cannell, D. S. \& G. Ahlers [1993]  "Spiral defect chaos in large aspect ratio {R}ayleigh-{B}\'enard convection"
\newblock \textit{Phys. Rev. Lett.} \textbf{71}, 2026-2029.

\noindent
Rajesh, S. \& Sinha, S. [2008]  "Measuring collective behavior of
  multicellular ensembles: role of space-time scales,"
\newblock \textit{J. Biosci} \textbf{33(2)}, 289--301.

\noindent
Shima, S. \& Kuramoto, Y. [2004]  "Rotating spiral wave with
  phase-randomized core in nonlocally coupled oscillators,"
\newblock \textit{Phys. Rev. E} \textbf{69}, 036213.

\noindent
Strogatz, S.H. [2001]  "Exploring complex networks,"
\newblock \textit{Nature} \textbf{410}, 268--276.

\noindent
Tanaka, D. \& Kuramoto, Y. [2003]  "Complex {G}inzburg-{L}andau equation
  with nonlocal coupling,"
\newblock \textit{Phys. Rev. E} \textbf{68}, 026219.

\noindent
Timme, M. [2006]  "Does dynamics reflect topology in directed networks,"
\newblock \textit{Europhys. Lett.} \textbf{76(3)}, 367--373.

\noindent
Vadivasova, T.E., Strelkova, G.I. \&  Anishchenko,V.S. [2001]
   "Phase-frequency synchronization in a chain of periodic oscillators
  in the presence of noise and harmonic forcings,"
\newblock \textit{Phys. Rev. E} \textbf{63}, 036225.

\noindent
Vanag, V.K., Yang, L., Dolnik, M., Zhabotinsky, A. M. \& Epstein, I.R. [2000]  "Oscillatory cluster
  patterns in a homogeneous chemical system with global feedback,"
\newblock \textit{Nature} \textbf{406}, 389--391.

\noindent
Wiesenfeld, K., Colet, P. \& Strogatz, S.H. [1996]  "Synchronization transitions in a disordered {J}osephson series array,"
\newblock \textit{Phys. Rev. Lett.} \textbf{76}, 404--407.

\noindent
Winfree, A.T. [1967]  "Biological rhythms and the behavior of
  populations of coupled oscillators,"
\newblock \textit{J. Theoret. Biol.} \textbf{16}, 15--42.

\noindent
Zhou, C. \& Kurths, J. [2006]  "Dynamical weights and enhanced
  synchronization in adaptive complex networks,"
\newblock \textit{Phys. Rev. Lett.} \textbf{96}, 164102.

\bigskip

\newpage

\end{document}